\documentstyle[12pt]{article}
\begin{document}
\def\be{\begin{equation}}
\def\en{\end{equation}}
\def\bear{\begin{eqnarray}}
\def\enar{\end{eqnarray}}
\def\beas{\begin{eqnarray*}}
\def\enas{\end{eqnarray*}}
\def\bera{ \setcounter{enumi}{\value{equation}} 
           \addtocounter{enumi}{1}
           \setcounter{equation}{0}
           \renewcommand{\theequation}{\theenumi\alph{equation}}
           \begin{eqnarray} }
\def\enra{ \end{eqnarray}
           \setcounter{equation}{\value{enumi}}
           \renewcommand{\theequation}{\arabic{equation}}}
\def\gsim{\stackrel{>}{\sim}}
\def\mn{{\mu\nu}}
\def\buildchar#1#2#3{\null \! \mathop {\vphantom {#1}
\smash #1}\limits ^{#2}_{#3}\!\null }
\def\ut#1{\buildchar{#1}{}{^\sim}\/}
\def\dsp{\displaystyle}
\def\ptl{\partial}
\def\ptlx{\partial_x}
\def\ptlt{\partial_t}
\def\itwo{\dsp{i\over 2}}
\def\im{\Im}
\def\re{\Re}
\newcommand{\half}{\frac{1}{2}}
\newcommand{\quat}{\frac{1}{4}}
\def\non{\nonumber \\}
\def\nonn{\nonumber \\ &&}
\def\dtri{\tilde{E}}
\def\ep{\epsilon}
\def\edth{\rlap/\partial}
\def\CA{{\cal A}}
\def\CB{{\cal B}}
\def\CC{{\cal C}}
\def\CD{{\cal D}}
\def\CI{{\cal I}}
\def\CJ{{\cal J}}
\def\den{e}
\renewcommand{\thefootnote}{\fnsymbol{footnote}}
\setcounter{footnote}{1}

\begin{flushright} 
 WU-AP/47/95  
(revised version)\\
 gr-qc/9602026 \\
 February 6, 1996   
\end{flushright}

\begin{center} 
\baselineskip .35in
\vskip 1.0cm
{\LARGE{\bf 
Constraints and Reality Conditions \\ in the 
Ashtekar Formulation \\ of General Relativity
}}
\vskip 1.0cm
\baselineskip .25in
{\large
{\sc Gen Yoneda}\footnote{Electronic address: 
yoneda@cfi.waseda.ac.jp}
and 
{\sc  Hisaaki Shinkai}\footnote{Electronic address: 
shinkai@cfi.waseda.ac.jp}   
} ~\\ 

$~{\mbox \dag}$ {\em Department of Mathematics, Waseda University,  \\
Okubo 3-4-1, Shinjuku, Tokyo 169, Japan} \\
$~{\mbox \ddag}$ {\em  Department of Physics, Waseda University, \\
 Okubo 3-4-1, Shinjuku, Tokyo 169, Japan} 

\vskip 1.5cm 
\end{center}
\begin{abstract}
\baselineskip .225in

We show how to treat the constraints and 
reality conditions in the $SO(3)$-ADM (Ashtekar) 
formulation of general relativity, 
for the case of a vacuum spacetime with a cosmological constant.
We clarify the difference between the reality conditions on  
the metric and on the triad. Assuming the triad
reality condition, we 
 find a new variable, allowing us to solve the gauge constraint
equations and the reality conditions simultaneously.

\vskip 0.5cm 
 \noindent
 Key Words: \parbox[t]{10.0cm}
 {General Relativity, Connection Formulation}

~

\noindent
To appear in {\it Classical and Quantum Gravity}

\end{abstract}
\vfill
\newpage
 \baselineskip .30in
\section{Introduction}

The $SO(3)$-ADM (Ashtekar) formalism of 
general relativity\cite{Ashtekar}
has many advantages in the treatment of gravity. 
In this formulation,
the constraint equations  are classified as 
the Hamiltonian constraint equation, ${\cal C}_H$, and
momentum constraint equations,  ${\cal C}_M$,  
with a new set of additional gauge constraint equations, ${\cal C}_G$.
These constraints become low-order 
polynomials and do not contain the inverses of the variables.
This formulation enables us to treat a quantum description of gravity
nonperturbatively. 

In order to apply the Ashtekar
formalism in classical general relativity, 
we need to solve the reality condition
for the metric and the extrinsic curvature.
In this paper, we show how to treat the constraints and 
reality conditions, 
for a vacuum spacetime with/without a cosmological constant. 
We stand on the point of pursuing the dynamics 
of spacetime, using evolutions of time-constant slices by fixing
gauge (slicing) condition in each time step. 
After a brief review of the Ashtekar formulation (\S 2), 
we clarify the difference between the reality conditions on  
the metric and on the triad (\S 3). We show that the latter condition
restricts a part of the gauge freedoms.

Assuming the triad reality condition, we 
find a new variable, which allows us to solve  ${\cal C}_G$
and the reality conditions simultaneously (\S 4). 
This technique is motivated by the works of 
Capovilla, Jacobson and Dell (CDJ) \cite{CDJ} and Barbero
\cite{barbero}.
CDJ discovered that in the vacuum spacetime 
the introduction of an arbitrary traceless and symmetric SO(3) tensor 
 makes two constraints ${\cal C}_H$ and ${\cal C}_M$ 
trivial, leaving only the third constraint ${\cal C}_G$
to be solved, while Barbero developed a technique for making 
${\cal C}_H$ and ${\cal C}_G$ trivial. 
Our new variable is analogous to both CDJ's and Barbero's,
but has advantage of clarifying
 the meaning of the additional constraint
${\cal C}_G$ in terms of ADM variables.


We use greek letters ($\mu, \nu, \rho, \cdots$), which range over 
the four spacetime coordinates $0, \cdots, 3$, 
while uppercase latin letters from the middle of
the alphabet ($I, J, K, \cdots$) range over the four internal 
SO(1,3) indices  $(0), \cdots, (3)$.
Lower case latin indices from the middle of the alphabet 
$(i, j, k, ...)$ range over the three spatial indices $1, \cdots, 3$,
while lower case latin indices from the beginning of the alphabet
$(a, b, c, ...)$ range over the three internal SO(3) indices
 $(1), \cdots, (3)$\footnote{We raise and lower 
$\mu,\nu,\rho$ by $g^{\mu\nu}$ and $g_{\mu\nu}$ (Lorenzian metric);
$I,J,K$ by $\eta^{IJ}={\rm diag}(-1,1,1,1)$ and $\eta_{IJ}$;
$i,j,k$ by $\gamma^{ij}$ and $\gamma_{ij}$(3-metric).
}.
We use volume forms
$\ep_{abc}$; $\ep_{abc} \ep^{abc}=3!$. 

\section{Brief review of the Ashtekar formulation}

The key feature of  Ashtekar's formulation of general relativity
\cite{Ashtekar} is the introduction of a self-dual 
connection as one of the basic dynamical variables.
Let us write the metric $g_\mn$ using the tetrad, $e^I_\mu$, and 
define its inverse, $E^\mu_I$, by 
$g_{\mu\nu}=e^I_\mu e^J_\nu \eta_{IJ}$ and 
$E^\mu_I:=e^J_\nu g^{\mu\nu}\eta_{IJ}$.
We define a SO(3,C) self-dual connection
\be \CA^a_{\mu} 
:= \omega^{0a}_\mu-\dsp{i \over 2} \ep^a_{~bc}\omega^{bc}_\mu,  
\label{w2A}
\en
where $\omega^{IJ}_{\mu}$ is a spin connection 1-form (Ricci 
connection), $\omega^{IJ}_{\mu}:=E^{I\nu} \nabla_\mu e^J_\nu.$
Note that the extrinsic curvature, 
$K_{ij}=-(\delta_i^{~l}+n_in^l)\nabla_ln_j$
in the ADM formalism, where $\nabla$ is a covariant 
derivative on $\Sigma$, 
satisfies the relation $-K_{ij}E^{ja}=\omega^{0a}_{i}$, 
when the gauge condition $E^0_a=0$ is fixed.
So 
$\CA^a_{i}$ is also expressed by
\be \CA^a_i = -K_{ij}E^{ja}
-\dsp{i \over 2} \ep^a_{~bc}\omega^{bc}_i. \label{def-ash}
\en

The lapse function, $N$, and shift vector, $N^i$,
are expressed as $E^\mu_0=({1 \over N}, -{N^i \over N}$). 
Ashtekar  treated the set  ($\CA^a_{i}$, $\dtri^i_{a}$) 
as basic dynamical variables, where 
$\dtri^i_{a}$ is an inverse of the densitized triad 
defined by $\dtri^i_{a}:=\den E^i_{a}$, and
where $\den:=\det e^a_i$ is a density.
This pair forms the canonical set
\bera
\{ \dtri^i_{a}(x), \dtri^j_{b}(y) \}
&=& 0, \label{poisson1}\\
\{ \CA^a_{~i}(x), \dtri^j_{b}(y) \}
&=& i \delta^j_{~i} \delta^a_{~b} \delta(x-y), \label{poisson2}\\
\{ \CA^a_{~i}(x),  \CA^b_{~j}(y) \}
&=&0.\label{poisson3}
\enra

The Hilbert action takes the form 
\be
S=\int {\rm d}^4 x 
[ \dot{\CA}^a_{i} \dtri^i_{a}
+\itwo \ut N \dtri^i_a \dtri^j_b F_{ij}^{c} \ep^{ab}_{~~c} 
-2\Lambda \ut N \det\dtri
-N^i F^a_{ij} \dtri^j_a  
+\CA^a_{0}\CD_i \dtri^i_{a} ], \label{action}
\en
where $\ut N := \den^{-1}N$, $\Lambda$ is cosmological constant, 
$\CD_i \dtri^i_{a}
    :=\ptl_i \dtri^i_{a}-i \ep_{ab}^{~~c}\CA^b_{i}\dtri^i_{c}$, and 
where ${F}^a_{\mu\nu}$ is curvature 2-form, defined as 
${F}^a_{\mu\nu} := \partial_\mu \CA^a_\nu - \partial_\nu \CA^a_\mu 
-\dsp{i\over 2} \ep^a_{~bc}(\CA^b 
 \wedge \CA^c)_{\mu\nu}$, and 
${\rm det}\dtri$ is defined to be  
${\rm det}\dtri=\dsp{1\over 6}
\ep^{abc}\ut\ep_{ijk}\dtri^i_a \dtri^j_b \dtri^k_c$, where
$\ep_{ijk}:=\ep_{abc}e^a_ie^b_je^c_k$
 and $\ut \ep_{ijk}:=\den^{-1}\ep_{ijk}$\footnote{$\ep_{xyz}=\den$, 
$\ut\ep_{xyz}=1$, $\ep^{xyz}=\den^{-1}$, $\tilde{\ep}^{xyz}=1$.}.

Varying the action with respect to
 the non-dynamical variables $\ut N$, $N^i$ 
and $\CA^a_{0}$ yields the constraint equations, 
\bera
{\cal C}_{H} &=& 
 \dsp{\ptl {\cal L} \over \ptl \ut N}=
 \itwo \ep^{ab}_{~~c} \dtri^i_{a} \dtri^j_{b} F_{ij}^{c} 
 -2\Lambda \det\dtri
  \approx 0, \label{c-ham} \\
{\cal C}_{M i} &=& 
  \dsp{\ptl{\cal L} \over \ptl N^i}=
  -F^a_{ij} \dtri^j_{a} 
\approx 0, \label{c-mom}\\
{\cal C}_{Ga} &=& \dsp{\ptl {\cal L} \over \ptl \CA^a_0}=
 \CD_i \dtri^i_{a}  
 \approx 0.  \label{c-g}
\enra
\noindent
The equations of motion for the dynamical variables
($\CA^a_i$ and $\dtri^i_a$) are 
\bera \dot{\CA}^a_{i} &=&
-i \ep^{ab}_{~~c}\ut N \dtri^j_{b} F_{ij}^{c}
+N^j F^a_{ji}
+\CD_i\CA^a_{0} 
+2\den\Lambda \ut N e^a_i,
\label{eqA} 
\\
\dot{\dtri^i_a}
&=&-i\CD_j( \ep^{cb}_{~~a} \ut N \dtri^j_{c}\dtri^i_{b})
+2\CD_j(N^{[j}\dtri^{i]}_{a})
+i\CA^b_{0} \ep_{ab}^{~~c} \dtri^i_c,  \label{eqE}
\enra
\noindent
where 
$\CD_jT^{ji}_a:=\ptl_jT^{ji}_a-i
 \ep^{a~c}_{~b} \CA^b_{j}T^{ji}_c,$ 
 for $T^{ij}_a+T^{ji}_a=0$.

\section{Reality conditions}
To ensure the metric is real-valued,  we need to impose two conditions;
the first is that the doubly densitized contravariant metric
$\tilde{\tilde \gamma}{}^{ij} := \den^2 \gamma^{ij}$ is real,
\bera
\Im (\dtri^i_a \dtri^{ja} ) = 0, \label{w-reality1} 
{\rm\bf ~~~~~~metric~reality~condition}
\enar
and the second condition is that the time derivative of
$\tilde{\tilde \gamma}{}^{ij}$ is real,
\bear
\Im \{ \ptl_t(\dtri^i_a \dtri^{ja} ) \} = 0.
{\rm\bf ~~~~~~second~metric~reality~condition}
 \label{w-reality2} 
\enra
\noindent
We denote these condition the ``metric reality condition" and the 
``second  
metric reality  condition" (extrinsic curvature reality condition), 
hereafter.  
Ashtekar {\it et al.} \cite{Ashtekar89a} discovered that, 
with the second metric reality condition (\ref{w-reality2}),
the reality of the 3-metric and extrinsic curvature
are automatically preserved under
time evolution, as a consequence of the equations of motion.
This means we need only solve both reality conditions 
(\ref{w-reality1})
and 
(\ref{w-reality2}) on the initial hypersurface.
Immirzi\cite{Immirzi} found that the reality conditions 
are consistent with the constraints, making the theory equivalent to 
Einstein's.

Using the equations of motion for $\dtri^i_{a}$ (\ref{eqE}), 
the gauge constraint (\ref{c-g}) and 
the first reality condition (\ref{w-reality1}), we can replace 
the second reality condition (\ref{w-reality2})
with a different constraint (see Appendix)
\be
W^{ij}:=\Re (\ep^{abc} 
\dtri^k_a \dtri^{(i}_b \CD_k \dtri^{j)}_c) \approx 0,  
\label{w-reality2-final}
\en
which fixes six components of $\CA^a_{i}$ and $\dtri^i_a$.
Moreover, in order to recover the original lapse function 
$N := \ut N \den$,
we demand $\Im (N/\den)=0$, i.e. the density $\den$ be real and
positive.
This requires that $\den^2$ be positive, i.e.
\bera
{\rm det}\dtri>0. \label{t-reality1}
\enar
Note that the metric reality conditions only 
guarantee the reality of $\den^4$.
The secondary condition of (\ref{t-reality1}),
\bear
\im[\ptl_t({\rm det}\dtri)]=0, \label{t-reality2}
\enra
\noindent
is automatically satisfied as a consequence of 
the equations of motion for $\dtri^i_{a}$ (\ref{eqE}),
the gauge constraint (\ref{c-g}),
the metric reality conditions (\ref{w-reality1}), (\ref{w-reality2})
and the first condition (\ref{t-reality1}) 
(see Appendix).
Therefore, in order to 
ensure that $\den$ is real, we only require (\ref{t-reality1}).
Note that this condition does not remove any
 degrees of freedom for the variables 
and is analogous to making the implicit assumption of
 $\det\gamma_{ij}> 0$ in the ADM formulation.

We now show that 
rather stronger reality conditions are useful 
in  Ashtekar's formalism for recovering the 
real 3-metric and extrinsic 
curvature.  These conditions are 
\bera
\Im (\dtri^i_a ) &=& 0  
{\rm\bf ~~~~~~first~triad~reality~condition}
\label{s-reality1} \\
{\rm and~~} 
\Im  ( \dot{\dtri^i_a} ) &=& 0, 
{\rm\bf ~~~~~second~triad~reality~condition}
\label{s-reality2} 
\enra
\noindent
and we denote them the ``first triad reality condition" and the 
``second triad
reality condition", hereafter.  
Using the equations of motion of $\dtri^i_{a}$, 
the gauge constraint (\ref{c-g}),
the metric reality conditions (\ref{w-reality1}), (\ref{w-reality2})
and the first condition (\ref{s-reality1}),
we see  (in Appendix) that  (\ref{s-reality2}) is equivalent to 
\be
\re(\CA^a_{0})=
\ptl_i( \ut N )\dtri^{ia}
+\half \den^{-1}e^b_i\ut N\dtri^{ja}\ptl_j\dtri^i_b
+N^{i}\Re(\CA^a_i). \label{s-reality2-final}
\en 
From this expression we see that 
the second triad reality condition 
restricts the three components of ``triad lapse" vector 
$\CA^a_{0}$\footnote{This ``triad lapse" is named by A.Ashtekar in
private communication.}.
Therefore (\ref{s-reality2-final}) is 
not a restriction on the dynamical variables ($\CA^a_i$ and $\dtri^i_a$)
but on a part of slicing, which we should impose on each hypersurface.
Thus the second triad reality condition does not restrict the 
dynamical variables any
further than the second metric condition does.

\section{Solving the constraint equations}
The equations we need to solve for $\CA^a_i$ and $\dtri^i_a$
are the constraints  (\ref{c-ham}), (\ref{c-mom}), (\ref{c-g})
and the reality conditions  
(\ref{w-reality2-final}), (\ref{s-reality1}).
CDJ solved ${\cal C}_H$ and ${\cal C}_M$
by introducing new variables. 
These reduced the 36 (real) independent components of
$\CA^a_i$ and $\dtri^i_a$ to 28, or  in CDJ's variables the 
 18 (real) independent components of $\psi_{ab}$ are
 reduced to 10 (a symmetric and traceless tensor), which 
corresponds to Weyl curvature $\Psi_i$.  
These  
are again restricted by ${\cal C}_G$ and the reality condition.

In contrast to CDJ's method, we make an alternative treatment of
the gauge constraint (\ref{c-g}) and the second metric 
reality condition 
(\ref{w-reality2-final}).
For convenience, we assume that $\dtri^i_a$ is real.
This assumption (\ref{s-reality1}) restricts our choice of triad, 
but this constraint is not difficult to satisfy.
We introduce 
the connection with double internal indices (note that here we do not 
use the densitized triad),
\be
\CA^{ab}:=\CA^a_i E^{ib}, \label{Aabteigi}
\en
and express all the constraints with 
($\CA^{ab}, \dtri^i_a$) as the basic pair of variables. 
The real part of ${\cal C}_G$ gives 
\[
\re({\cal C}^a_G)=
\ptl_i\dtri^{ia}+ \ep^{a~c}_{~b} \im(\CA^b_i)\dtri^i_c
=
\ptl_i\dtri^{ia}+\den \ep^a_{~bc}\im(\CA^{bc})=0,
\]
where $\den = \sqrt{\det \dtri}$. 
Thus the imaginary and anti-symmetric part of $\CA^{ab}$
is determined from 
\be
\im(\CA^{[ab]})
=-\dsp{1 \over 2\den}\ep^{abc}\ptl_i\dtri^i_c.
\label{Aia}
\en
The imaginary part of ${\cal C}_G$ gives 
\[
\im({\cal C}^a_G)=
-\ep^{a~c}_{~b} \re(\CA^b_i)\dtri^i_c
=-\den \ep^a_{~bc} \re(\CA^{bc})=0.
\]
Thus the real and anti-symmetric part of $\CA^{ab}$ is 
\be
\re(\CA^{[ab]})=0.\label{Ara}
\en
Thus we have confirmed that the 6 real constraints of ${\cal C}_G$ 
are automatically satisfied  if 
we impose (\ref{Aia}) and (\ref{Ara}).

Next the second metric reality condition (\ref{w-reality2-final}) 
becomes 
\[
W^{ij}=
\half[
\epsilon^{abc} \dtri^k_a \dtri^i_b \ptl_k\dtri^j_c
+\epsilon^{abc} \dtri^k_a \dtri^j_b \ptl_k\dtri^i_c
+2 \dtri^i_b \dtri^{jb} \im({\CA^a}_a) 
-2\dtri^i_b \dtri^j_a \im(\CA^{(ba)})]=0. 
\]
Thus the imaginary and symmetric part of $\CA^{ab}$ is 
\bear
\im(\CA^{(ab)})=
\dsp{1\over 2}[
E^j_d(\ep^{dac} \ut e^b_i +\epsilon^{dbc}\ut e^a_i)
\ptl_j \dtri^i_c 
-\delta^{ab}
E^j_d  \ep^{d~c}_{~e} \ut e^e_i \ptl_j \dtri^i_c ],
\label{Ais}
\enar
where $\ut e^a_i$ is the inverse of $\dtri^i_a$. 
 From these expressions, 
we see that
(\ref{c-g}) and (\ref{w-reality2-final}) are satisfied if and only if
$\CA^{ab}$ satisfies (\ref{Aia}), (\ref{Ara}) and (\ref{Ais}),
after assuming (\ref{s-reality1}). 
We note that the imaginary part of $\CA^{ab}$ 
consists of the triad and its spatial differential
and that the real part of $\CA^{ab}$ is symmetric.


These results become clearer if we compare $\re(\CA^{ab})$
and the extrinsic curvature $K_{ij}$ through the definition of 
$\CA^a_i$, (\ref{def-ash}).  
 From (\ref{def-ash}) we derive
\bera
\re(\CA^{ab})&=&-K_{ij}E^{ia} E^{jb},           \label{remind1}\\
\im(\CA^{ab})&=&-\half \ep^a_{~cd}\omega^{cd}_i E^{ib}. 
\label{remind2}
\enra
\noindent
Since the extrinsic curvature is symmetric, we see $\re(\CA^{ab})$ is
also symmetric [(\ref{remind1})].
After some calculation, we can see that (\ref{remind2}) 
is equivalent to (\ref{Aia}) and (\ref{Ais}).
Moreover,  we find from  (\ref{def-ash}) that 
\[
K_{ij}=-\CA^a_i e_{ja}-\itwo \ep_{abc}\omega^{bc}_i e^a_j,
\]
so the reality of the extrinsic curvature, $\im (K_{ij})=0$,
is equivalent to (\ref{remind2}). 
Consequently, 
$\re(\CC_G)=0$ [(\ref{Aia})] and  $W^{ij}=0$ [(\ref{Ais})]
indicates that the extrinsic curvature is real 
and 
$\im(\CC_G)=0$ [(\ref{Ara})] 
indicates that the extrinsic curvature is symmetric.


When one has solved the 12 equations ${\cal C}_G=0$ and $W^{ij}=0$
for the 27 variables $\CA^a_i$(complex) and $\dtri^i_a$(real),
 15 degrees of freedom remain.
Introducing $\CA^{ab}$ clarifies this remaining freedom;
these are 6 degrees of freedom for  $\re(\CA^{(ab)})$ and 
9 for $\dtri^i_a$.
Our task is now reduced to solving the 
other constraints (\ref{c-ham})  and  (\ref{c-mom}) 
for the variables $\re(\CA^{(ab)})$ and $\dtri^i_a$.

 In terms of $\re(\CA^{(ab)})$ and $\dtri^i_a$, the constraints
are  given by substituting  (\ref{Aia}), (\ref{Ara}) and (\ref{Ais})
 into
(\ref{c-ham})  and  (\ref{c-mom}). Then we see  $\im({\cal C}_H)=0$ 
and 
$\im({\cal C}_M)=0$ are automatically satisfied \cite{Immirzi},
thus the equations which we need to solve are just four equations;
\bera
\re({\cal C}_H)&=& \den [
  \ep_{ab}^{~~c}  \dtri^i_c (\ptl_i I^{ab})
+\half \ep_{ab}^{~~c}   \dtri^i_c \ut e^d_j  \ptl_i\dtri^j_d I^{ab}
+ \ep_a^{~cd}\dtri^i_c \ut e_{jb} \ptl_i \dtri^j_d I^{ab}
\non&&
+\dsp{1\over 2} (R^2
-  I^2
-  R^{ab}R_{ba}
+  I^{ab}I_{ba})
-2\Lambda ] \approx 0, \label{finalCH}
\\
\re({\cal C}_{Mi})
&=& \den[
- \ptl_i R 
+ \ut e_{ia}\dtri^j_b \ptl_j R^{ba}
-\half   \ut e ^d_j (\ptl_i\dtri^j_d) R 
+ \ut e_{jc} (\ptl_i \dtri^j_b)R^{bc}
\non &&
+\half   \dtri^k_b \ut e_{ia}  \ut e^d_j(\ptl_k\dtri^j_d)R^{ba}
- \dtri^k_b \ut e^a_i \ut e_{jc} (\ptl_k\dtri^j_a)R^{bc} 
-\ep_{bcd}\den  R^{ab}I^{cd} \ut e_{ia}] \approx 0
\label{finalCM}
\enra
where $R^{ab}=\re(\CA^{ab})=\re(\CA^{(ab)})$, $I^{ab}=\im(\CA^{ab})$,
$R={R^a}_a$ and $I={I^a}_a$.
These are equivalent to the scalar and vector constraints
in ADM formulation. 

The Poisson bracket for this pair of the variables becomes 
\bera
\{ \dtri^i_{a}(x), \dtri^j_{b}(y) \}
&=& 0, \label{poisson11}\\
\{ \CA^{ab}(x),\dtri^i_c(y)\}&=&
i{1\over \den} \dtri^{ib}(x)\delta^a_c\delta(x-y), 
\label{poisson22}\\
\{ \CA^{ab}(x),  \CA^{cd}(y) \}
&=&0.\label{poisson33}
\enra
We expect that our variables are convenient for expressing
the data on each hypersurface if we impose a reality condition.
However, we remark that, like CDJ's variable, our variables 
($\re(\CA^{(ab)})$, $\dtri^i_a$) are not canonical. 
Therefore, when we describe the equations of motion of our variables, 
we transform those of canonical pair
[e.g., (\ref{eqA}) and (\ref{eqE})] into ours. 
Also note that the formulation is not polynomial. 
Consequently, our full set of equations consists of four constraint
equations (\ref{finalCH}), (\ref{finalCM}) [together with 
definitions (\ref{Aia}), (\ref{Ara}) and (\ref{Ais})], 
and the equations of motion (\ref{eqA}), (\ref{eqE}).

\section{Discussion}

We have studied the $SO(3)$-Ashtekar formulation 
from the point of pursuing the dynamics
of spacetime, using evolutions of time-constant slices. 
We examined the difference between the reality conditions on  
the metric and on the triad, and demonstrated that the latter 
condition
restricts a part of the gauge freedoms [$\re(\CA^a_0)$]. 
When we apply this condition in time evolution problems (based on 3+1
decompositions), this restriction of gauge variables must be imposed 
at every time step.
Having assumed the triad reality condition, we 
find a new variable, allowing us to solve ${\cal C}_G$
and the reality conditions simultaneously. 
Our variable clarifies the meanings of the additional constraint
and the second reality condition,
which express the reality and symmetry of the extrinsic curvature.

Let us now compare our variables and CDJ's.
CDJ's $\psi_{ab}$ is expressed by the Weyl scalar $\Psi_i$ in the 
Newman-Penrose formulation \cite{PR}, so that $\psi_{ab}$ has the 
same true degrees of freedom as the gravitational curvature.
While our variable  $\CA^{ab}$ is a ``connection", using it 
 instead of $\CA^a_i$ shows us the physical significance 
 of the gauge constraints and the 
second reality condition.
In CDJ's method, 
the remaining tasks are to solve 
the gauge constraints and the reality conditions
for $\CA^a_i$ and $\Psi_{ab}$,  
note that this is not simple in the CDJ variables.
In our method, we have only 4 equations to solve, as opposed to 
21 equations in CDJ's.

A practical application of this variable is expected to arise 
in numerical relativity. 
Recently, Salisbury {\it et al} \cite{Salisbury} proposed the use of 
CDJ formulation to numerical relativity, in which they expect 
to improve the boundary conditions for gravitational waves. 
We are now preparing a new approach to numerical treatment of gravity,
by combining our connection formulation together with the ADM, Ashtekar
 and CDJ formulations to express data on the 3-hypersurfaces. 
We expect that we can give new procedures
in evolving data, fixing slicing conditions and/or including
gauge field. Such a formulation and simulations 
will be presented elsewhere \cite{YSnumerical}.


We thank R. Easther for a careful reading of  our manuscript.
This work was supported partially by the Grant-in-Aid for Scientific
Research Fund of the Ministry of Education, Science, Sports and 
Culture No. 07854014 and by a Waseda University Grant for 
Special Research Projects.

\newpage
\section*{Appendix ~~~ 
 Details of the reality conditions} \label{sec:appc}
\renewcommand{\theequation}{A.\arabic{equation}}
\setcounter{equation}{0}

In this appendix, we derive 
the second metric reality condition (\ref{w-reality2-final}) and
the second triad reality condition (\ref{s-reality2-final}),
and show that $\im[\ptl_t({\rm det}\dtri)]=0$ is automatically 
satisfied.

First we derive
the second metric reality condition (\ref{w-reality2-final}).
We start from its original definition (\ref{w-reality2}),
\[
\im\{\ptl_t(\dtri^i_a\dtri^{ja})\}
=
2\im(\dot{\dtri^{(i}_a} \dtri^{j)a}).
\]
Using the equation of motion (\ref{eqE}) and 
 gauge constraint (\ref{c-g}), we have
\beas
\dot{\dtri^i_a} \dtri^{ja}
&=&
[-i\CD_k(\ep^{cba} \ut N \dtri^k_c \dtri^i_b)
+\CD_k(N^k\dtri^{ia})
-\CD_k(N^i\dtri^{ka})
-i\CA^b_{~0} \ep_b^{~ac} \dtri^i_c]\dtri^j_a
\non &=&
[-i\ep^{cba}\CD_k(\ut N) \dtri^k_c \dtri^i_b
 -i\ep^{cba}\ut N \dtri^k_c \CD_k(\dtri^i_b)
+\CD_k(N^k)\dtri^{ia} 
\nonn
+N^k \CD_k(\dtri^{ia})
-\CD_k(N^i)\dtri^{ka}
-i\CA^b_{~0} \ep_b^{~ac} \dtri^i_c]\dtri^j_a, 
\enas
giving
\[
\dot{\dtri^{(i}}_a  \dtri^{j)a}
=
-i\ep^{cba}\ut N \dtri^k_c \CD_k(\dtri^{(i}_b)\dtri^{j)}_a
+\ptl_k(N^k)\dtri^i_a\dtri^{ja}
+N^k \CD_k(\dtri^{(i}_a)\dtri^{j)a}
-\dtri^k_a\ptl_k(N^{(i})\dtri^{j)a}. 
\]
Thus we obtain
\beas
\im(\dot{\dtri^{(i}_a} \dtri^{j)a})
&=&
 -\re[\ep^{cba}\ut N \dtri^k_c \CD_k(\dtri^{(i}_b)\dtri^{j)}_a]
+\half\im[N^k \ptl_k(\dtri^i_a\dtri^{ja})]
\\ &=&
 -\ut N \ep^{cba}\re[ \dtri^k_c \CD_k(\dtri^{(i}_b)\dtri^{j)}_a],
\enas
where we use the metric first reality condition (\ref{w-reality1}).
The vanishing of this gives (\ref{w-reality2-final}). $\Box$

Second, we show that
$\im[\ptl_t({\rm det}\dtri)]=0$ is automatically satisfied
when we assume the first density reality condition ${\rm det}\dtri>0$.
We have
\beas
\ptl_t({\rm det}\dtri)
&=&
\dsp{1\over 2}\ut \ep_{ijk}\ep^{ade}\dtri^j_d\dtri^k_e\dot{\dtri^i_a}\\
&=&
\dsp{1\over 2}\ut \ep_{ijk}\ep^{ade}\dtri^j_d\dtri^k_e
[
 -i\ep^{cb}_{~~a}\CD_l(\ut N) \dtri^l_c \dtri^i_b
 -i\ep^{cb}_{~~a}\ut N \dtri^l_c \CD_l(\dtri^i_b)
\nonn
+\CD_l(N^l)\dtri^i_a
+N^l \CD_l(\dtri^i_a)
-\CD_l(N^i)\dtri^l_a
-i\CA^b_{~0} \ep_{ba}^{~~c} \dtri^i_c] \\
 &=&
\dsp{1\over 2}\ut \ep_{ijk}\ep^{ade}\dtri^j_d\dtri^k_e
[ -i \ep^{cb}_{~~a}\ut N \dtri^l_c \CD_l(\dtri^i_b)
+\ptl_l(N^l)\dtri^i_a
+N^l \CD_l(\dtri^i_a)
-\ptl_l(N^i)\dtri^l_a] 
\enas
where we use the gauge constraint again.
The first term becomes
\beas
-i\dsp{1\over 2}\ut\ep_{ijk}\ep^{ade}\dtri^j_d\dtri^k_e
 \ep^{cb}_{~~a}\ut N \dtri^l_c \CD_l(\dtri^i_b)
&=&
-\dsp{i\over \den}\ep_{ijk}\dtri^{jc}\dtri^{kb}
 \ut N \dtri^l_c \CD_l(\dtri^i_b)
\\ &=&
i \den \ut N \ep_{ik}^{~~l} \dtri^{kb} \CD_l\dtri^i_b
\\ &=&
i \den \ut N \ep_c^{~be} e^c_i \dtri^l_e  \CD_l\dtri^i_b.
\enas
Now we have 
\[
W^{ij}\gamma_{ij}
=
\ep^{abc}\re(\dtri^k_a\dtri^i_b\CD_k\dtri^j_c)e^d_i e_{jd}
=
\den \ep^{a~c}_{~b} \re(\dtri^k_a e^b_j \CD_k\dtri^j_c)
=
\den \ep^{e~c}_{~b} \re(\dtri^l_e e^c_i \CD_l\dtri^i_b).
\]
Since this vanishes by the second metric reality condition,
we see that the imaginary part of the first term is zero.
Thus we have
\beas
\Im[\ptl_t({\rm det}\dtri)]
&=&
\Im\{ \dsp{1\over 2}\ut \ep_{ijk}\ep^{ade}\dtri^j_d\dtri^k_e
[\ptl_l(N^l)\dtri^i_a
+N^l \CD_l(\dtri^i_a)
-\ptl_l(N^i)\dtri^l_a] \}
\non &=&
\Im[
3\den^2 \ptl_l(N^l)
+N^l \ptl_l(\den^2)
-\den^2 \ptl_l(N^l)]
=0, 
\enas
where we use the assumption ${\rm det}\dtri=\den^2>0$.
Thus the second condition is automatically satisfied. $\Box$

Next we show the second triad reality condition is written in the
form of (\ref{s-reality2-final})
when we assume 
the second metric reality condition 
(\ref{w-reality2}) and 
the first triad reality condition $\im(\dtri^i_a)=0$. 
Since $\dtri^i_a$ is non degenerate, there exists $P_{ab}$ such that
\[
\im(\dot{\dtri^i_a})=P_{ab}\dtri^{ib}.
\]
Using the metric second reality and the first triad reality,
we have
\[
0=\im(\dot{\dtri^{(i}_a})\dtri^{j)a}
={P_a}^b \dtri^{(i}_b\dtri^{j)a}
=P_{(ab)}\dtri^{ib}\dtri^{ja}
\]
which implies $P_{(ab)}=0$.
Thus the second triad reality conditions is equivalent to 
$P_{[ab]}=0$.
Let us derive $P_{ab}$.
\beas
P_{ab}&=& 
\den^{-1}\Im(\dot{\dtri^i_a})e_{ib}
\\&=&
\den^{-1}e_{ib}
\Im[-i\CD_j(\ep^{cd}_{~~a} \ut N \dtri^j_{c}\dtri^i_{d})
+2\CD_j(N^{[j}\dtri^{i]}_{a})
+i\CA^d_{0}\ep_{ad}^{~~c} \dtri^i_c]
\\&=&
\den^{-1}e_{ib}
\Im[
-i\ep^{cd}_{~~a}\ptl_j( \ut N )\dtri^j_{c}\dtri^i_{d}
-i\ep^{cd}_{~~a}\ut N \dtri^j_{c}
 (\ptl_j\dtri^i_{d}-i\ep_{de}^{~~f} \CA^e_j\dtri^i_f)
+\ptl_j(N^{j})\dtri^{i}_{a}
\\&&
+N^{j} (\ptl_j\dtri^i_{a}-i \ep_{ae}^{~~f} \CA^e_j\dtri^i_f)
-\ptl_j(N^{i})\dtri^{j}_{a}
+i\CA^d_{0} \ep_{ad}^{~~c} \dtri^i_c]
\\&=&
\den^{-1}e_{ib}[
-\ep^{cd}_{~~a}\ptl_j( \ut N )\dtri^j_{c}\dtri^i_{d}
-\ep^{cd}_{~~a}\ut N \dtri^j_{c}\ptl_j\dtri^i_{d}
-\ep^{cd}_{~~a}\ut N \dtri^j_{c} \ep_{de}^{~~f} \Im(\CA^e_j)\dtri^i_f
\\&&
-N^{j} \ep_{ae}^{~~f}\Re(\CA^e_j)\dtri^i_f
+\Re(\CA^d_{0}) \ep_{ad}^{~~c} \dtri^i_c]
\\&=&
-\ep^c_{~ba}\ptl_j( \ut N )\dtri^j_{c}
-\den^{-1}e_{ib} \ep^{cd}_{~~a}\ut N \dtri^j_{c}\ptl_j\dtri^i_{d}
-\ut N  \dtri^j_{b}\Im(\CA_{ja})
+\ut N  \dtri^j_{c}\Im(\CA^c_j)\delta^{ab}
\\&&
-N^{j} \ep_{aeb}\Re(\CA^e_j)
+\Re(\CA^d_{0})\ep_{adb}.
\enas
Thus $P_{[ab]}=0$ becomes 
\beas
\ep^{abc}P_{ab}
&=&
\ep^{abc}
[
-\ep^d_{~ba}\ptl_j( \ut N )\dtri^j_{d}
-\den^{-1}e_{ib} \ep^{ed}_{~~a}\ut N \dtri^j_{e}\ptl_j\dtri^i_{d}
-\ut N  \dtri^j_{b}\Im(\CA_{ja})
+\ut N  \dtri^j_{d}\Im(\CA^d_j)\delta^{ab}
\\&&
-N^{j} \ep_{aeb}\Re(\CA^e_j)
+\Re(\CA^d_{0})\ep_{adb}
]
\\&=&
2\ptl_j( \ut N )\dtri^{jc}
-\den^{-1}e^b_i\ut N\dtri^j_b\ptl_j\dtri^{ic}
+\den^{-1}e^b_i\ut N\dtri^{jc}\ptl_j\dtri^i_b
-\ep^{abc}\ut N  \dtri^j_{b}\Im(\CA_{ja})
\\&&
+2N^{j}\Re(\CA^c_j)
-2\Re(\CA^c_{0})
\\&=&
2\ptl_j( \ut N )\dtri^{jc}
-\ut N \Re(\CD_i\dtri^{ic})
+\den^{-1}e^b_i\ut N\dtri^{jc}\ptl_j\dtri^i_b
+2N^{j}\Re(\CA^c_j)
-2\Re(\CA^c_{0})
\\&=&
2\ptl_j( \ut N )\dtri^{jc}
+\den^{-1}e^b_i\ut N\dtri^{jc}\ptl_j\dtri^i_b
+2N^{j}\Re(\CA^c_j)
-2\Re(\CA^c_{0})
\enas
This last equation vanishes, giving (\ref{s-reality2-final}). $\Box$

~

~

~

\end{document}